\documentclass[twocolumn,trackchanges]{aastex631}

\usepackage{orcidlink}

\begin{document}
\title{Observational Evidence for Magnetic Field Amplification in SN 1006 }

\author{Moeri Tao \orcidlink{0009-0008-9030-760X}}
\affiliation{School of Advanced Science and Engineering, Wseda University, 3-4-1 Okubo, Shinjuku-ku, Tokyo, Japan}

\author{Jun Kataoka$^*$ \orcidlink{0000-0003-2819-6415}}
\affiliation{School of Advanced Science and Engineering, Wseda University, 3-4-1 Okubo, Shinjuku-ku, Tokyo, Japan}
\footnote{$^*$Corresponding Author: kataoka.jun@waseda.jp}

\author{Takaaki Tanaka \orcidlink{0000-0002-4383-0368}}
\affiliation{Department of Physics, Konan University, 8-9-1 Okamoto, Higashinada, Kobe, Hyogo, Japan}

\begin{abstract}
We report the first observational evidence for magnetic field amplification in the north-east/south-west (NE/SW) shells of supernova remnant SN 1006, one of the most 
promising sites of cosmic ray (CR) acceleration.
In previous studies, the strength of magnetic fields in these shells was estimated to be $B_{\rm SED}$ 
$\simeq$ 25$\mu$G from the 
spectral energy distribution, where the synchrotron emission from relativistic electrons accounted for radio to X-rays, along with the inverse Compton emission extending from the GeV to TeV energy bands. However, the analysis of broadband radio data, ranging from 1.37~GHz to 100~GHz, indicated that the radio spectrum steepened from $\alpha_1 = 0.52 \pm 0.02$ to $\alpha_2 = 1.34 \pm 0.21$ by $\Delta \alpha$ = 0.85 $\pm$ 0.21. This is naturally interpreted as a cooling break under strong magnetic field of $B_{\rm brk}$ $\ge$ 2~mG. Moreover, the high-resolution MeerKAT image indicated that the width of the radio NE/SW shells was broader than that of the X-ray shell by a factor of only 3$-$20, as measured by Chandra. Such narrow radio shells can be naturally explained if the 
magnetic field responsible for the radio emissions is $B_{\rm R}$ $\ge$ 2 mG.  Assuming that the magnetic field is locally enhanced by a factor of approximately $a$ = 100 along the NE/SW shells, we argue that the filling factor, which is the volume ratio of such a magnetically enhanced region to that of the entire shell, must be as low as approximately $k$ = 2.5$\times$10$^{-5}$.
\end{abstract}

\keywords{acceleration of particles - ISM: individual objects(SN 1006) - ISM: magnetic fields - ISM: supernova remnants - radio continuum: ISM}

\section{Introduction} \label{sec:intro}
Young supernova remnants (SNRs) are considered to be promising sites for the production of galactic cosmic ray (CRs) through diffusive shock acceleration \citep[DSA;][]{bell1978acceleration,blandford1987particle,malkov2001nonlinear}, 
although the detailed process of DSA is not well understood. In particular, the magnetic field strength and structure are vital for determining the maximum energy of particles that can be accelerated in the shock.

SN~1006, located at a distance of 1.8~kpc \citep{green2001}, is a prototypical example wherein synchrotron X-ray emissions along the outer north-east (NE) and south-west (SW) shells are detected through observation from the ASCA. Assuming an estimated magnetic field of $B$ $\simeq$ 6--10~\rm{$\mu$}G, \cite{koyama1995evidence} found that electrons of $\ge$100~TeV were being accelerated in the shock of SN 1006. High-resolution Chandra Advanced CCD Imaging Spectrometer (ACIS) images revealed that the X-ray shells were extremely thin with scale widths of 4\arcsec (0.04 pc) and 20\arcsec (0.2 pc) in the upstream and downstream regions, respectively \citep{bamba2003small}. The authors assumed downstream and upstream magnetic field strengths of $B_d$ = 4$B_u$ = 40 $\mu$G, although larger values, namely, $B_d$ $\simeq$ 150 $\mu$G, are suggested with field amplification \citep{ksenofontov2005dependence}.

Very high energy gamma rays (above 100 GeV) from SN~1006 were detected by the High Energy Stereoscopic System (H.E.S.S.) (\cite{acero2010first}). Together with the GeV gamma-ray observation from the  Fermi-Large Area Telescope (LAT) over 10 years observations, \cite{xing2019fermi} suggested that the spectral energy distribution (SED) was well represented by a one-zone leptonic model, wherein synchrotron emission constituted the low energy bump from radio to X-rays, whereas inverse Compton (IC) emission on cosmic microwave background (CMB) accounts for the high energy bump in GeV/TeV gamma rays. Based on SED modeling, the magnetic field strengths were estimated to be $B_{\rm SED}$ $\simeq$ 24~$\mu$G and 30~$\mu$G for the NE and SW shells, respectively. However, recent Planck observations indicated a spectral curvature above 10~GHz (\cite{arnaud2016planck}); thus, the radio-to-X-ray spectrum may not connect smoothly as opposed to that determined by SED models by various authors. This is supported by the fact that the optical/ultra-violet (UV) counterpart of SN~1006 is extremely faint except for the bright H$\alpha$ filament in the NW shell \citep{winkler2003sn,korreck2004far}.  

Comparisons with similar young SNRs may provide some hints to solve the contradiction regarding the magnetic field strengths. For example, the overall SED of RX~J1713.7$-$3946 is well represented by the synchrotron and IC (CMB) model with $B$ $\simeq$ 10 $\mu$G, as in the case of SN~1006 (but see \cite{ellison2010efficient} to account for the overall spectrum of RX~J1713.7$-$3946 with a hadronic model). However, short time variability on a one-year timescale was found in the shell of  RX~J1713.7$-$3946; thus, the magnetic field in the hot spot should be as high as $B_{\rm HS}$ $\simeq$ 1~mG (\cite{uchiyama2007extremely}).  Similar to that of RX~J1713.7$-$3946, Cassiopia A indicates very thin non-thermal X-ray fillament along the shock, and a fast variability timescale of 4 years was found, which also suggests the amplification of the magnetic field to 
$B_{\rm HS}$ $\simeq$ 1~mG (\cite{uchiyama2008fast}). 

In this Letter, we present new and independent evidence for magnetic field amplification in the NE/SW shells of SN 1006. In particular, we did not rely on detecting  such short time variability but focused on the spectral and imaging results.  We presented the systematic analysis of radio data from 1.4 to 100~GHz to confirm the spectral break in the radio spectrum. We also estimated the optical and UV fluxes to fill in the SED gap between radio and X-rays. Finally, we compared the high-resolution radio and X-ray image, which provides independent evidence for an enhanced magnetic field in the NE/SW shell of SN 1006, and then subsequently presented the conclusion of  our findings.

\section{Analysis and result} \label{sec:style}
\subsection{Broadband Radio Spectrum}
We analyzed all the archival Planck-low frequency instrument (LFI) data containing 
30, 44, and 70 GHz fluxes, and 100~GHz data from high frequency instrument (HFI) to identify the high frequency spectral features of SN~1006.
Based on \cite{arnaud2016planck}, the flux densities were measured using standard aperture photometry, with the source size centered on SN~1006 and the diameter scaled to 
$\theta_{\rm ap}$ = 1.5 $\theta_s$.
Here, $\theta_s$ is the source size estimated from 
$\theta_s$ = $\sqrt{\theta_{\rm SNR}^2 + \theta_{b}^2}$, where $\theta_{\rm SNR}$ and $\theta_b$
are the spatial distribution of SN~1006 and the Planck beam size \citep{ade2014planckb,aghanim2014planckb}, respectively.  $\theta_{\rm SNR}$ is 15\arcmin 
 ~and  $\theta_b$ are 16.2\arcmin~at 30 GHz, 13.6\arcmin~at 44 GHz and 6.7\arcmin~at 70GHz, and 4.83\arcmin~at 100 GHz. We also estimated and 
subtracted the background from the area between concentric circles with inner and outer radii of 1.5 $\theta_{\rm ap}$ and 2.0 $\theta_{\rm ap}$, respectively. An aperture correction was applied to correct for the loss of flux density outside the aperture. The uncertainties for the flux densities were the root-sum-square of calibration uncertainty \citep{ade2014planckc,aghanim2014planckc} and propagated statistical errors. The statistical errors considered the number of pixels in the source aperture and background annulus, and the standard deviation within the background annulus \citep{arnaud2016planck}.

\begin{figure}[h]
    \centering
    \includegraphics[scale=0.65]{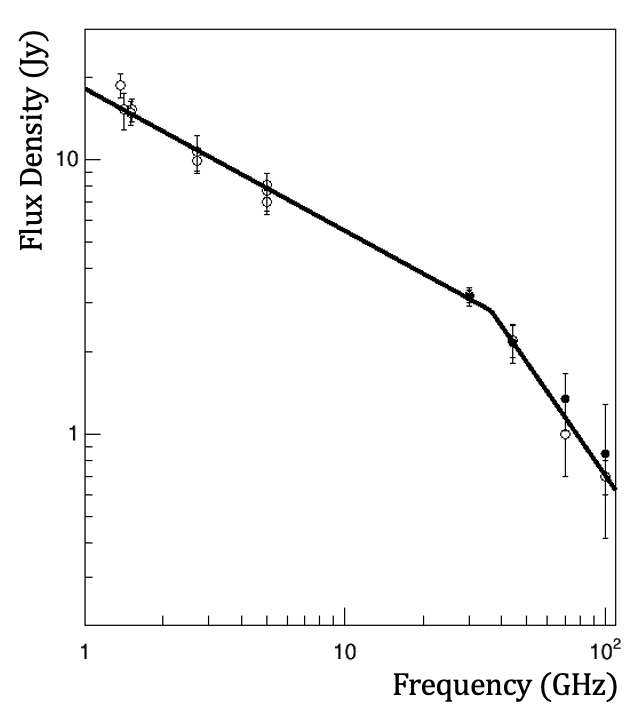}
    \caption{The broadband radio spectrum of SN 1006. The $filled$ $circles$ are derived from Planck measurements in this study. The $open$ $circles$ represent the archival radio flux densities from \cite{dyer20091}, \cite{petruk2009aspect}, \cite{milne1971supernova}, \cite{rothenflug2004geometry}, \cite{gardner1965supernova}, \cite{kundu1970brightness}, \cite{milne19755}, and \cite{arnaud2016planck}.}
    \label{fig:SN1006radio}
\end{figure}

Figure \ref{fig:SN1006radio} shows the broadband radio spectrum of SN 1006 as measured with Planck ($filled$ $circles$), along with a collection of previous measurements ranging from 1.4 to 100~GHz ($open$ $circles$). 
The spectrum cannot be represented by a single power law of $S_\nu \propto \nu^{-\alpha}$, where $S_\nu$ is the flux density at frequency $\nu$. 
Rather, it can be well represented by a 
broken power law with a break frequency of
$\nu_{brk}=36 \pm 6\rm{GHz}$, above which the spectral index steepens from $\alpha_1 = 0.52 \pm 0.02$ to $\alpha_2 = 1.34 \pm 0.21$, where  the reduced chi-squared for the fitting is $\chi^2$/dof = 6.74/13. The confidence level that the broken power law is a better representation of 
the data than a single power law is 99.9$\%$ (3.59 $\sigma$) based on the $F$-test. 

\begin{figure*}[t]
    \begin{center}
    \includegraphics[scale=0.5]{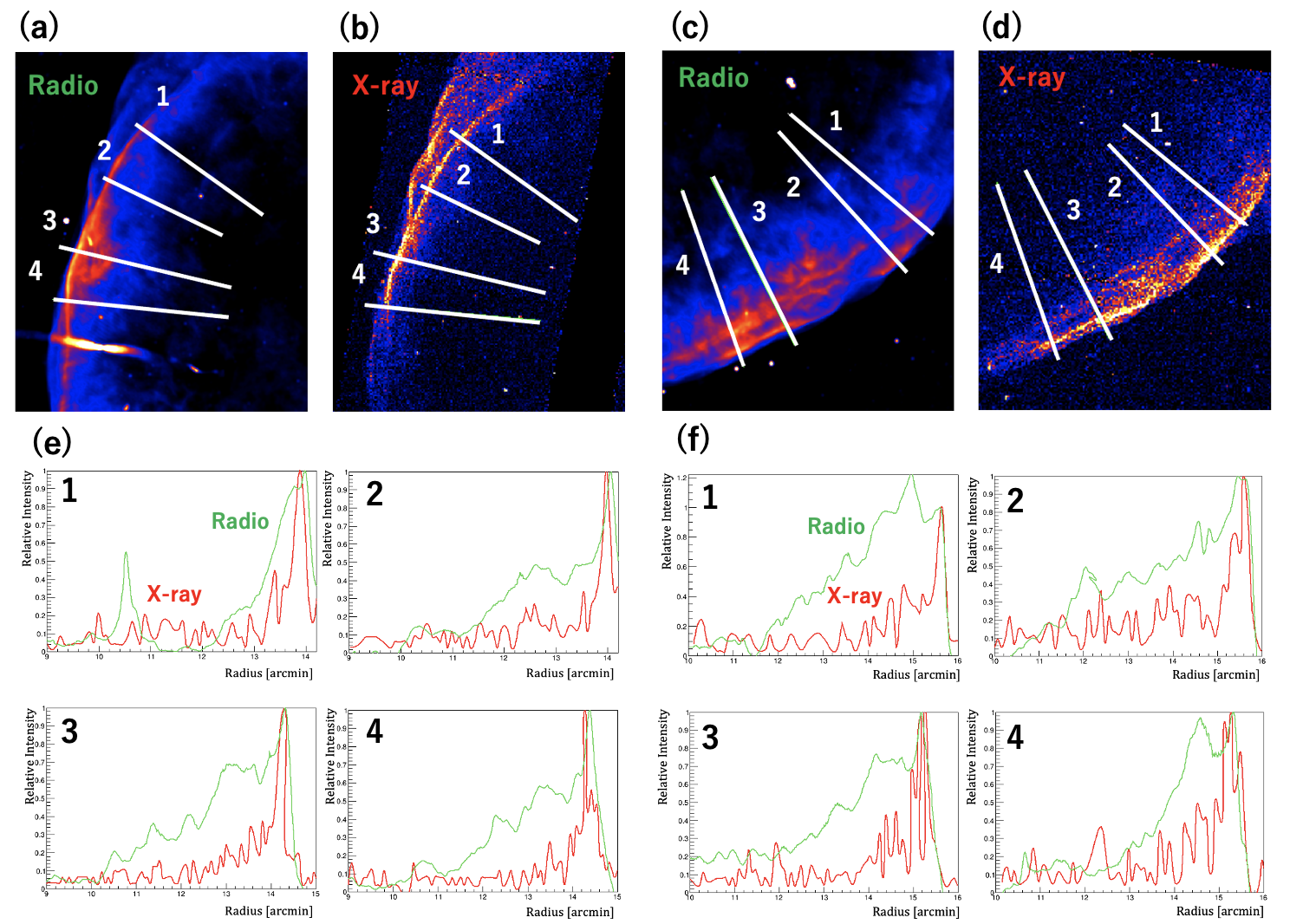}
    \caption{$Upper$: Comparison between the MeerKAT (1335 MHz) and  Chandra (2.0$-$7.0~keV) images of SN~1006. (a) MeerKAT and (b) Chandra images of the NE shell.  (c) MeerKAT and (d) Chandra images of the SW shell.  $Lower$: Comparison between the radial profiles of MeerKAT (1335 MHz) ($green$ line) and  Chandra (2.0$-$7.0~keV) ($red$ line) images across the (e) NE and (f) SW shells along the $white$ line. Note that, the spike structure seen in (a), that is vertical to the shock front is not related to SN1006 but the background FR-I radio galaxy, WISEA J150403.60-415550.9. Full details are given in p.24 of \cite{cotton2024meerkat}. Also in (f)-1, the radio profile is normalized at the second peak, which is more likely the shock front.}
    \label{fig:radialprofile}
    \end{center}
\end{figure*}
\subsection{Estimation of the Optical and UV Fluxes} \label{subsec:opuv}

The optical emission of SN 1006 is extremely faint except for the bright H$\alpha$ filament as observed in the NW shell \citep{winkler2003sn}. 
Thus, the only  H$\alpha$ filament has been extensively studied in terms of 
proper motion, as well as comparison with the shock structure in optical and X-rays. 
Compared with the maximum surface brightness for the filament, namely, $2.0\times 10^{-16} ~\rm{erg~cm^{-2}~s^{-1}~arcsec^{-2}}$, the intensity of the diffuse emission associated with NE/SW shells of SN~1006 is fainter tham a factor of 20$-$25 (see Fig.3 of \citealt{winkler2003sn}). Assuming a region size of 70 arcmin$^2$ for the NE shell, the integrated flux would be $\simeq$ $ (1.6\textrm{--}2.5) \times 10^{-12}~\rm{erg~cm^{-2}~s^{-1}}$ . Considering the entire SN~1006, we simply doubled the NE flux assuming that a similar amount of flux would by contributed by the SW shell.

The UV flux in SN~1006 is also faint, as measured by the Far Ultraviolet 
Spectroscopic Explorer (FUSE). The flux density in the NE shell in the UV range is approximately  $(0.1 \textrm{--} 0.2) \times 10^{-14}~\rm{erg~cm^{-2}~s^{-1}~\AA^{-1}}$ 
between 1010 and 1050 \AA\ (see Fig.2 of \citealt{korreck2004far}).
Considering the relatively narrow field of view (30\arcsec $\times$ 30\arcsec) of FUSE, the integrated NE flux is estimated to be $(1.1-2.2) \times 10^{-11}~ \rm{erg~cm^{-2}~s^{-1}}$ for the assumed region size of 70 arcmin$^2$. We simply doubled the NE flux to  estimate the flux from entire region of SN~1006.

\subsection{Direct comparison of radio/X-ray shells}

Although SN~1006 is a prototypical SNR that 
clearly exhibits non-thermal shells/filaments in both radio and X-rays, direct comparisons have not been made until now using high resolution radio and X-ray images with angular resolutions better than scale width 
($\sim$ 10\arcsec) of the intrinsic shell width. Figure \ref{fig:radialprofile} shows the radial profiles  of NE/SW shells of SN 1006, as 
compared with those of the radio (1335 MHz; MeerKAT) and X-rays (2.0$-$7.0~keV ; Chandra). Comparisons 
were made across various parts of the shells along the $white$ lines. MeerKAT is a radio telescope in Northern Cape province of South Africa \citep{jonas2009meerkat}, whose angular resolution at 1335 MHz is $\simeq$~8\arcsec \citep{cotton2024meerkat}. This is still worse than the X-ray image resolution provided by Chandra ACIS ($\simeq$~0\farcs5) but better than that of the previous image provided by the Very Large Array (e.g., $\simeq$ ~20\arcsec; \citealt{reynolds1986radio}). 
Figure \ref{fig:radialprofile} clearly shows 
that the radio shell is broader than that of X-rays, but at most, it is  only approximately 10-fold of both the NE and SW shells.

\newpage

\section{Discussion and Conclusion} \label{sec:floats}
\subsection{Estimation of the magnetic field strength}
In $\S$2.1, we reported that the broadband radio spectrum showed a spectral break at 
$\nu_{\rm brk}$ = 36 $\pm$ 6~GHz, at which the spectrum steepened from $\alpha_1 = 0.52 \pm 0.02$ to $\alpha_2 = 1.34 \pm 0.21$ by $\Delta \alpha$ = 0.85 $\pm$ 0.21. This result has confirmed the finding of \cite{arnaud2016planck}. Notably, that such a break in the spectral index is often seen as a result of synchrotron losses in various active galactic nuclei (AGNs) \citep[e.g.,][]{inoue1996electron,kataoka1999high}.
Assuming that this is a cooling break at which the electron-cooling time $t_{\rm cool}$ and dynamical timescale of SNR $t_{\rm {dyn}}$ are balanced, we expect $t_{\rm {cool}}$ $\simeq$ $t_{\rm {dyn}}$ at $\nu$ = $\nu_{\rm{brk}}$. For the synchrotron emission, the cooling time of electrons can be expressed as  $t_{\rm {cool}}\simeq 5.1\times 10^8 {B_{\rm{cool}}^{-2}\gamma_{\rm brk}^{-1}}$, where $\gamma_{\rm {brk}}$ is the Lorentz factor of electrons emitting $\nu_{\rm {brk}}$ photons in the magnetic field $B_{\rm {cool}}$, and  $\nu_{\rm{brk}} \simeq 1.2 \times 10^6$ $B_{\rm {cool}} {\gamma_{\rm {brk}}}^2$ \citep[e.g.,][]{Rybicki:847173}. Subsequently, we can obtain 
\begin{equation}
    B_{\rm brk} \lbrack\rm{\mu G}\rbrack \simeq 6.8 \times 10^3 {\it t_{\rm cool}\,{\lbrack \rm{kyr} \rbrack}}^{-\frac{2}{3}} {\nu_{brk}\,\lbrack  \rm{GHz} \rbrack}^ {-\frac{1}{3}}.
\end{equation}
The dynamical timescale of SNR, $t_{\rm dyn}$, 
can be regarded as the advection time of electrons across
the shock, which should be shorter than the age of the SNR. Substituting the age of SN~1006, namely,  
$t_{\rm age}$ $\simeq$ 1 kyr,  
for $t_{\rm cool}$, we can obtain a lower limit of $B_{\rm brk}$ $\ge$ 2~mG for $\nu_{brk}$ = 36 GHz.
This is {\it the magnetic field estimated from the ``cooling break" in the radio spectrum}. In this formulation, only the synchrotron cooling is considered. 
The contribution from the cooling of IC (CMB) cooling is more than one order of magnitude smaller and can be ignored as we descibed below.

Another way of estimating the magnetic field from the observed radio flux is to assume an equipartition (i.e., minimum energy) between the electron and magnetic field energy densities, $U_e$ = $U_B$. This can be expressed as   
\begin{equation}
B_{\rm min} [\mu \rm G] = 27\Bigl( {\frac{\eta\hspace{1mm} {\it d}[\rm kpc]^2 {\it f}_\nu [\rm Jy]}{{\it V} [\rm pc^3]} \Bigr)^{2/7}},
\end{equation}
where $d$ and $V$ are the distance and volume of the SNR, and 
$f_{\nu}$ is the flux density at frequency $\nu$ (eg. \cite{Longair}).
$\eta$ is the ratio of the energy stored in electrons and protons, where $\eta$ = 1 for $e^-$-$e^+$ plasma. 
Consequently, we obtained $B_{\rm min}$ $\simeq$  5 $\mu$G. Instead, if we assume a typical CR composition,  $\eta$ $\simeq$ 100, 
$B_{\rm min}$ $\simeq$ 10 $\mu$G is obtained.
Hence, {\it the magnetic field estimated from the equipartition} is an order of 
10$\mu$G regardless of the plasma content; thus,  
$B_{\rm min}$ $\ll$ $B_{\rm brk}$.

\subsection{Implication from SED}
\begin{figure}[ht]
    \centering
    \includegraphics[scale = 0.59]{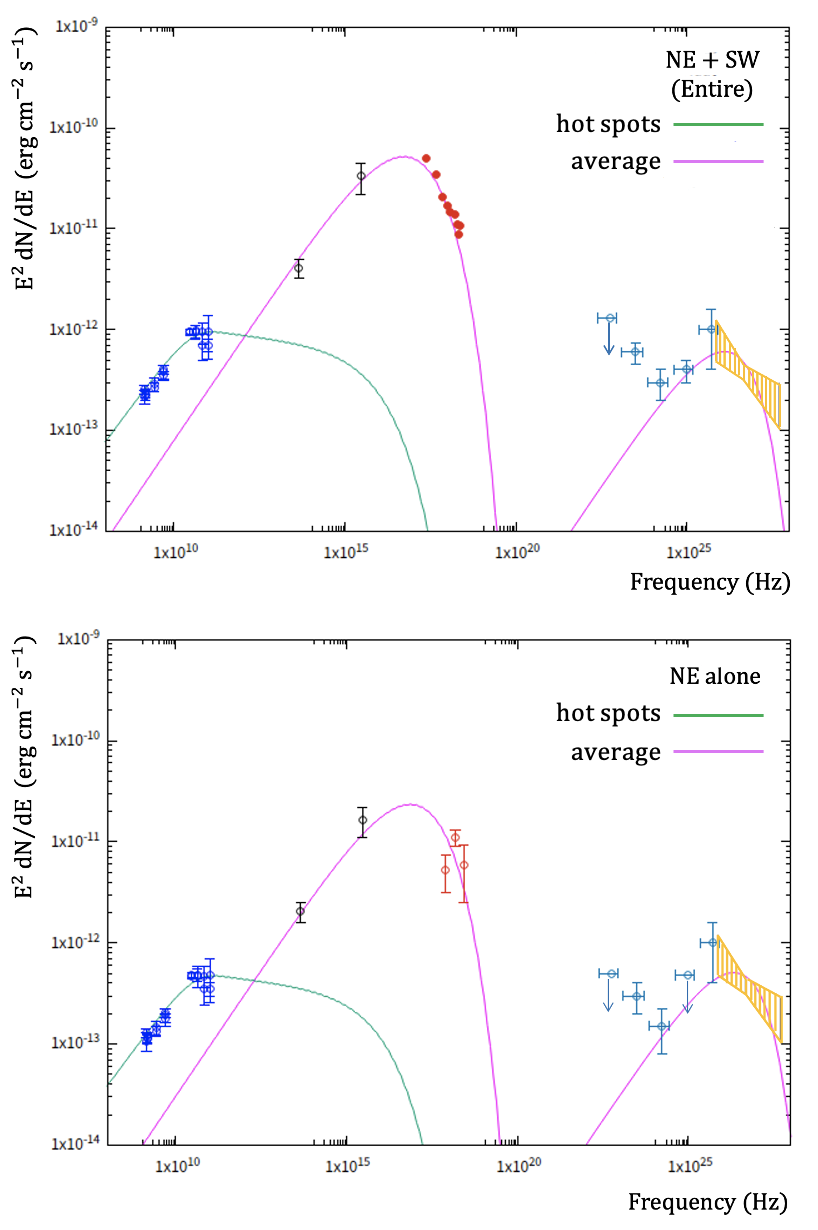}
    \caption{$Top$: Double synchrotron and IC (CMB) model fit to the broadband SED of the entire SN 1006 (including both NE+SW shells). Radio (this study; $blue$ $open$ $circles$), Suzaku X-ray (\cite{bamba2008suzaku}; $filled$ $red$ $circles$), Fermi (\cite{xing2019fermi}; $cyan$ $open$ $circles$) data, and H.E.S.S. (\cite{acero2010first}; $yellow$ $region$) data are shown with the first synchrotron 
     ($green$ $dashes$), second synchrotron and its 
     IC (CMB) ($magenta$ $dashes$) emissions, respectively. $Black$ $open$ $circles$ represent the doubled values of the optical and UV fluxes obtained for the NE shell.
     $Bottom$: same as above, but the SED is fit to the NE shell of SN 1006. $Blue$ $open$ $circles$ represent half the value of the radio fluxes obtained for the entire remnant.
     The $red$ $open$ $circles$ represent the X-ray measurement for the NE  shell reported in \cite{kalemci2006x}. The optical (\cite{winkler2003sn}) and ultraviolet (\cite{korreck2004far}) data are represented as $black$ $open$ $circles$ .}
    \label{fig:SN1006SED}
\end{figure}

Subsequently, we construct the SED from radio to TeV gamma rays, 
including the broadband radio spectrum 
and optical/UV fluxes newly derived in this study. Figure \ref{fig:SN1006SED} shows the $updated$ broadband SED of the entire SN~1006 (NE + SW shells) and that of the NE shell alone. In previous studies, the overall  SED was modeled by one-zone synchrotron/IC(CMB) emissions with s moderate magnetic field of $B_{\rm SED}$ $\simeq 25\mu$G \citep[e.g.,][]{xing2019fermi}. 
However, according to our results, multiple components are necessary to account for the radio-to-X-ray spectrum. Notably, the first synchrotron component dominates the radio emissions
but flattens above $\nu_{\rm brk}$. The second 
synchrotron component appears weak in the radio 
band but accounts for most of the optical-UV-X-ray 
emission without turnover. 

Thus, we assumed ``double" electron populations to reproduce the overall structure of the SED;  the first population is responsible for the 
synchrotron emission in compact regions or hot spots where the magnetic field is enhanced by a factor of $a$ $\simeq$ 100; thus, $B_{\rm HS}$ $\simeq$ $B_{\rm brk}$ $\simeq$ $a$$\times$$B_{\rm min}$.  Conversely, the second population radiates 
another synchrotron emission in the ``average" magnetic 
field of $B_{\rm SED}\hspace{1mm} $ 
$\simeq$ $B_{\rm min}$. 
Although the geometry of the magnetic enhanced 
regions is still unknown, the similarity of the radio and X-ray images, as shown in Figure \ref{fig:radialprofile}, may imply that such regions are almost uniformly distributed along the shock as either compact patches, like hot spots, or as diffuse sheets/filaments. Hereafter, we call such 
magnetically enhanced regions ``hot spots" for simplicity. In any case, we can constrain the filling factor $k$, which is the volume ratio of such magnetically enhanced regions to the entire shell, from the observed SED. The radio flux in the hot spots
is expressed as  
$f_{\rm HS}$ $\propto$ $U_{e,\rm{HS}}$ $U_{B,\rm {HS}}$ $k$ $V$. 
Assuming that the radio emissions of hot spots is 10--100 times brighter than that of the other (i.e., average) parts of the SNR, $f$ $\propto$ $U_{e}$$U_{B}$(1$-$$k$)V, we obtain $f_{\rm H}/f$ $\simeq$ $k$$\times$$a^3$ $\sim$ 10--100; thus, 10$^{-5}$ $\le$ $k$ $\le$ 10$^{-4}$. 

An example of SED fits is also shown in  Figure \ref{fig:SN1006SED} assuming
a filling factor of $k$ = 2.5$ \times $10$^{-5}$. 
The $magenta$ represents the synchrotron/IC (CMB) emission from entire (average) region, whereas the $green$ curve is the synchrotron emission in the hot spots. Notably, the corresponding IC (CMB) emission in the hot spot is too low to appear in this figure. The first electron population has a cooling break at $\gamma_{\rm{brk}}$ = 2$ \times $10$^3$ (or $\gamma_{\rm brk}$$m_e c^2$ = 1~GeV), above which the electron spectral index $s$ steepened from 2.1 to 3.1.
We assumed that $B_{\rm{HS}}$ = 2.5~mG in the hot spots to be consistent with the observed break at $\nu_{\rm{brk}}$ = 36 GHz.  In contrast, the second population has no break in the electron spectrum with $s$ = 2.0 up to the maximum 
frequency of $\gamma_{\rm {max}}$ = 1.0$\times$$10^8$ (or  $\gamma_{\rm max}$$m_e c^2$ = 51~TeV). The magnetic fields were assumed to be $B_{\rm {SED}}$ = 25 $ \mu $G  and 18 $\mu$G for the entire SN~1006 and NE shell, respectively. 
Table \ref{tab:SED} summarizes the fitting parameters of the SEDs. 
Note that $B_{\rm SED}$ is well determined within an uncertainty of $\simeq$30~$\%$, 
 whereas $B_{\rm HS}$ = 2.5~mG, is just an example of possible fitting parameters. Similarly,  
 $\gamma_{\rm max}$ = 1.0
 $\times$ 10$^6$ for the hot spots is an assumption because it cannot be determined solely from the SED. 

\setlength\intextsep{0pt}
\begin{table}[]
    \caption{The fitting parameters of the SEDs}
\begin{flushleft}
    \begin{tabular}{ccc}
   \hline
    Parameter & Entire (hot spots) & ~~Entire (average)\\
    \hline
    $\rm{ N_0 ~[ cm^{-3} \gamma^{-1}]} $&$ 4.5 \times 10^{-5} $&$ 7.5 \times 10^{-8} $  \\
    $ \gamma_{\rm{min}}$ & 1 & 1\\
    $\gamma_{\rm{brk}}$ & $2.0 \times 10^3$ & $-$\\
    $ \gamma_{\rm{max}} $&$ 1.0 \times 10^6 $& $ 1.0 \times 10^8 $ \\
    index ~$s$& 2.1 &  2.0\\
     $B$ [$\mu$G] & $ 2500  $&$ 25 $\\
     volume $V$ [pc$^3$] & $ 8.5 \times 10^{-2} $&$ 3.5 \times 10^3 $\\
    \hline
     parameter &NE (hot spots)  &NE (average) \\
    \hline
    $\rm{ N_0 ~ [ cm^{-3} \gamma^{-1}]} $& $ 4.5 \times 10^{-5} $ &$ 1.5 \times 10^{-6} $ \\
    $ \gamma_{\rm{min}}$  &1&1\\
    $\gamma_{\rm{brk}}$ & $2.0 \times 10^3$&$-$\\
    $ \gamma_{\rm{max} }$&$ 1.0 \times 10^6 $&$ 1.0 \times 10^8 $\\
    index~ $s$& 2.1&2.0\\
    $B$ [$\mu$G] &$ 2500 $&$ 18 $\\
     volume $V$ [pc$^3$] & $ 4.2 \times 10^{-2}  $&$ 1.1 \times 10^3 $\\
    \hline
    \end{tabular}
    \vskip
    \baselineskip
    \small{Notes: As for an electron 
    population in the hot spots,  
    we assumed a broken power law function with an exponential cut-off, $N_e (\gamma) = N_0 \gamma^{-s} \rm{exp}(-\gamma/\gamma_{\rm max})$ for $\gamma_{\rm min} < \gamma < \gamma_{\rm brk}$, whereas  
    $N_e (\gamma) = N_0 \gamma_{\rm brk} \gamma^{-(s+1)} \rm{exp}(-\gamma/\gamma_{\rm max})$ for $\gamma > \gamma_{\rm brk}$. For remaining average region, we assumed a simple power law function with an exponential cutoff, $N_e (\gamma) = N_0 \gamma^{-s} \rm{exp}(-\gamma/\gamma_{\rm max})$.}
 \end{flushleft}
    \label{tab:SED}
\end{table}

\subsection{Comparison between Radio/X-ray of the NE/SW shells}
\begin{figure}
    \centering
    \includegraphics[scale = 0.43]{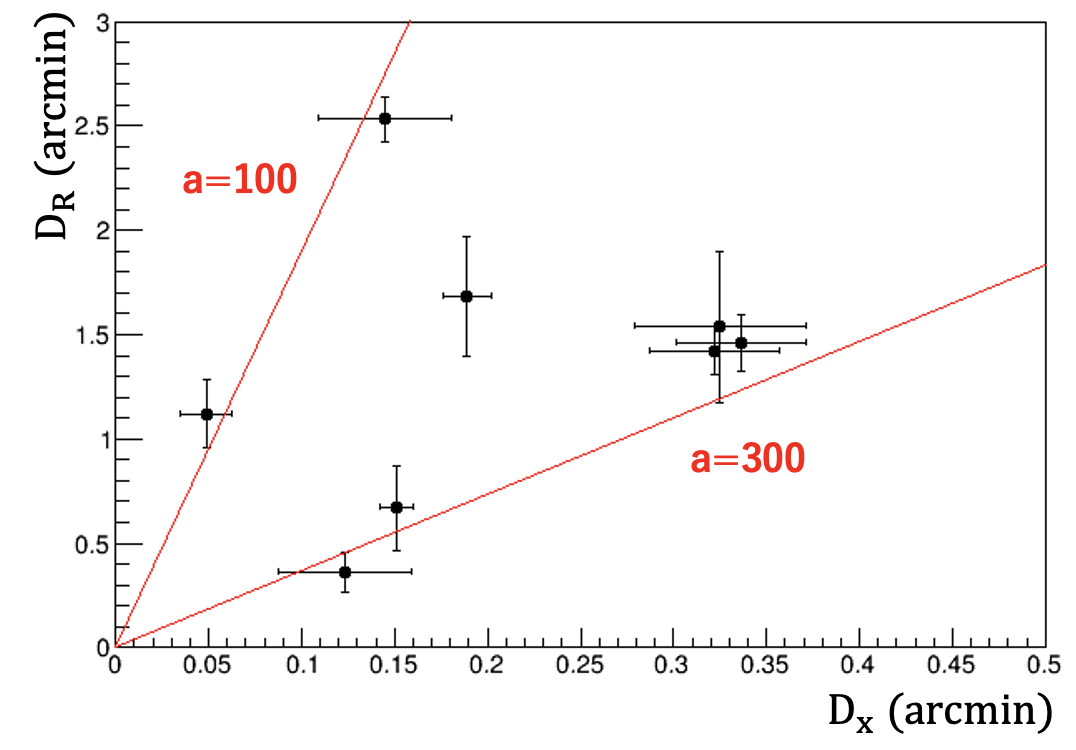}
    \caption{Comparison of thickness of the X-ray shell ($D_{\rm X}$) and radio  shell ($D_{\rm{R}}$) derived from 
    Figure \ref{fig:radialprofile}. The $red$ lines represent  
    $a$ = 100 and 300, which correspond to $D_{\rm X}$/$D_{\rm{R}}$ = 19.0 and 3.7, respectively.}
    \label{fig:widthratio}
\end{figure}

Subsequently, we discuss the spatial extent of the NE/SW shells as observed with the radio and X-ray images shown in Figure \ref{fig:radialprofile}. In general, the thickness of the shell can roughly be estimated as $D$ $\simeq$ $v_{\rm sh}$$\times$$t_{\rm cool}$, where $v_{\rm sh}$ is the shock speed and $t_{\rm cool}$ is the cooling time of electrons that emit photons of certain energies. Therefore, if the magnetic field for radio and X-ray emissions are 
the same, the thickness of the shell measured in radio ($D_R$) and X-rays ($D_X$) reflects the difference of $t_{\rm cool}$; thus, 
$\propto$ $\nu^{-1/2}$.  
Hence, we expect $D_R/D_X$ $\simeq$ 2$\times$10$^4$, which is, however,  far from what is shown in  Figure \ref{fig:radialprofile}.  In fact,  the width of NE/SW shells, defined here as the full width at half maximum (FWHM) of the peak differs by a factor of 3$-$20, as shown in Figure \ref{fig:widthratio}. Moreover, this difference is 
not owing to the different angular resolutions between Chandra and 
MeerKAT, as the width of the NE/SW shells is generally broader than 0\farcm4
in the radio band, that is more than three times that of the MeerKAT angular resolution.

The observed $D_{\rm{R}}/D_{\rm{X}}$ also indicates that the magnetic fields, which 
mainly contribute to the observed radio and X-ray emissions,  
cannot be the same. In such a case, 
\begin{equation}
    \frac{D_{\rm{R}}}{D_{\rm{X}}} = \Bigl( \frac{B_{\rm{R}}}{B_{\rm{X}}} \Bigr)^{-3/2} \Bigl( \frac{\nu_{\rm{R}}}{\nu_{\rm{X}}} \Bigr)^{-1/2},
\end{equation}
where $B_{\rm {R}}$ and $B_{\rm {X}}$ are the magnetic fields responsible for the radio and X-ray emissions. The observed ratio of $3 \leq D_{\rm{R}}/D_{\rm{X}} \leq 23$, can 
be converted to the amplification factor of the magnetic field 
as $88 \leq a \leq 348$.
This is consistent with magnetic field amplification in the hot spots, namely, $a$ $\simeq$ 100, which is independently estimated from the SED. 

Finally, assuming  $B_{\rm R}$ = $B_{\rm HS}$ $\simeq$ 2.5 mG and $B_{\rm X}$ = $B_{\rm SED}$  $\simeq$ 25 $\mu$G, the thickness 
of the shell is estimated as approximately
$D_{\rm R}$ $\simeq$ 3.7 pc (or 7\farcm2) 
and $D_{\rm X}$ $\simeq$ 0.2 pc (or 0\farcm4) 
for $v_{\rm sh}$ $\simeq$ 10$^3$ km/s.  
This is slightly larger, but almost consistent with that observed in the NE/SW shells as shown in Figures \ref{fig:radialprofile} 
and \ref{fig:widthratio}.

\subsection{Comparison with other SNRs}
The manifestation of hot spots with a strong magnetic field of $B_{\rm {HS}}$ of approximately 1~mG has been reported in similar young SNRs such as RX~J1713.7$-$3946 \citep{uchiyama2007extremely} and Cassiopeia~A \citep{uchiyama2008fast}. 
In both objects, the flickering of X-ray hot spots on yearly scales was observed. Meanwhile, the overall SED of 
 RX~J1713.7$-$3946 is well represented by the synchrotron/IC (CMB) model with  $B_{\rm SED}$ of $\sim$ 10 $\mu$G 
 \citep{abdo2011observations}, although even larger $B_{\rm SED}$ of 
$\simeq$ 250 $\mu$G is suggested for the case of Cassiopeia~A 
\citep{saha2014origin}. The filling factor of such 
hot spots, namely, $k$, is not discussed in \cite{uchiyama2007extremely}, but they have suggested that the X-ray emissions from the 
hot spots is less than approximately 1/100 of the entire emission 
of the SNR, which is consistent with the present observation of SN~1006 (see Figure \ref{fig:SN1006SED}). Note, however, that 
the anticipated linear size of the hot spots 
in SN~1006 
would be approximately $k^{1/3}$ $=$ 2.9$\times$10$^{-2}$
times smaller than the typical shell thickness 
of approximately 10\arcsec. Thus this is difficult to resolve, 
even with Chandra.

The physical mechanism of magnetic amplification 
in  the hot spots remains quite uncertain in all the SNRs. In the standard theory of shock compression, $a$ = $B_d$/$B_u$  = $U_{d}$/ $U_{u}$ = 4, where $B_d$ and $B_u$ are the downstream and upstream magnetic fields and $U_{d}$ and $U_{u}$ are the  downstream and upstream energy densities of the plasma, respectively.  Thus $a$ $\simeq$ 100 is hardly explained by the classical shock theory. Instead,  such a strong magnetic field may be produced owing to the turbulent dynamo action through shock-cloud interactions 
as suggested for RX~J1713.7$-$3946
\citep{inoue2011toward};  however, this cannot be the case for a relatively ``clean" environment as in SN~1006 and 
Cassiopeia A (but see, \citealt{miceli2014shock}). In this context, some particle-in-cell (PIC) simulations of non-relativistic perpendicular shocks in the high-Mach-number ($M_A$)
suggests magnetic amplification of $a$ = 
5.5($\sqrt{M_A} - 2$); thus $a$ $\simeq$ 100 for $M_A$ $\simeq$ 400 or $v_{\rm sh}$ $\simeq$ 
4000 km s$^{-1}$
(\cite{bohdan2021magnetic}). However, whether similar efficient amplification is possible even in parallel shocks is uncertain as observed in SN~1006 \citep{reynoso2013radio,zhou2023magnetic}.
Similar analysis using Planck,  MeerKAT, and Chandra is ongoing for other types of SNRs to systematically understand the physical origin of 
magnetic amplification.

\begin{acknowledgments}

We thank the anonymous referee 
for his/her helpful comments to 
improve the manuscript.
This research was supported by Japan Science and Technology Agency (JST) ERATO Grant Number JPMJER2102, Japan.
The Planck data and MeerKAT data are available from NASA LAMBA (\dataset[doi:10.26131/IRSA485]{https://doi.org/10.26131/IRSA485}) and the SARAO archive and image products  (\dataset[doi:10.48479/nz0n-p845]{https://doi.org/10.48479/nz0n-p845}).
This paper also employs a list of Chandra datasets, obtained by the Chandra X-ray Observatory, contained in~\dataset[doi:10.25574]{https://doi.org/10.25574/cdc.253}.
\end{acknowledgments}

\bibliography{SN1006_ApJL_fin}{}

\begin{thebibliography}{}
\expandafter\ifx\csname natexlab\endcsname\relax\def\natexlab#1{#1}\fi
\providecommand{\url}[1]{\href{#1}{#1}}
\providecommand{\dodoi}[1]{doi:~\href{http://doi.org/#1}{\nolinkurl{#1}}}
\providecommand{\doeprint}[1]{\href{http://ascl.net/#1}{\nolinkurl{http://ascl.net/#1}}}
\providecommand{\doarXiv}[1]{\href{https://arxiv.org/abs/#1}{\nolinkurl{https://arxiv.org/abs/#1}}}

\bibitem[{Abdo {et~al.}(2011)Abdo, Ackermann, Ajello, Allafort, Baldini, Ballet, Barbiellini, Baring, Bastieri, Bellazzini, {et~al.}}]{abdo2011observations}
Abdo, A., Ackermann, M., Ajello, M., {et~al.} 2011, The Astrophysical Journal, 734, 28

\bibitem[{Acero {et~al.}(2010)Acero, Aharonian, Akhperjanian, Anton, De~Almeida, Bazer-Bachi, Becherini, Behera, Beilicke, Bernl{\"o}hr, {et~al.}}]{acero2010first}
Acero, F., Aharonian, F., Akhperjanian, A., {et~al.} 2010, Astronomy \& Astrophysics, 516, A62

\bibitem[{Ade {et~al.}(2014{\natexlab{a}})Ade, Aghanim, Armitage-Caplan, Arnaud, Ashdown, Atrio-Barandela, Aumont, Baccigalupi, Banday, Barreiro, {et~al.}}]{ade2014planckb}
Ade, P., Aghanim, N., Armitage-Caplan, C., {et~al.} 2014{\natexlab{a}}, Astronomy \& Astrophysics, 571, A7

\bibitem[{Ade {et~al.}(2014{\natexlab{b}})Ade, Aghanim, Armitage-Caplan, Arnaud, Ashdown, Atrio-Barandela, Aumont, Baccigalupi, Banday, Barreiro, {et~al.}}]{ade2014planckc}
---. 2014{\natexlab{b}}, Astronomy \& Astrophysics, 571, A8

\bibitem[{Aghanim {et~al.}(2014{\natexlab{a}})Aghanim, Armitage-Caplan, Arnaud, Ashdown, Atrio-Barandela, Aumont, Baccigalupi, Banday, Barreiro, Battaner, {et~al.}}]{aghanim2014planckb}
Aghanim, N., Armitage-Caplan, C., Arnaud, M., {et~al.} 2014{\natexlab{a}}, Astronomy \& Astrophysics, 571, A4

\bibitem[{Aghanim {et~al.}(2014{\natexlab{b}})Aghanim, Armitage-Caplan, Arnaud, Ashdown, Atrio-Barandela, Aumont, Baccigalupi, Banday, Barreiro, Battaner, {et~al.}}]{aghanim2014planckc}
---. 2014{\natexlab{b}}, Astronomy \& Astrophysics, 571, A5

\bibitem[{Arnaud {et~al.}(2016)Arnaud, Ashdown, Atrio-Barandela, Aumont, Baccigalupi, Banday, Barreiro, Battaner, Benabed, Benoit-L{\'e}vy, {et~al.}}]{arnaud2016planck}
Arnaud, M., Ashdown, M., Atrio-Barandela, F., {et~al.} 2016, Astronomy \& Astrophysics, 586, A134

\bibitem[{Bamba {et~al.}(2003)Bamba, Yamazaki, Ueno, \& Koyama}]{bamba2003small}
Bamba, A., Yamazaki, R., Ueno, M., \& Koyama, K. 2003, The Astrophysical Journal, 589, 827

\bibitem[{Bamba {et~al.}(2008)Bamba, Fukazawa, Hiraga, Hughes, Katagiri, Kokubun, Koyama, Miyata, Mizuno, Mori, {et~al.}}]{bamba2008suzaku}
Bamba, A., Fukazawa, Y., Hiraga, J.~S., {et~al.} 2008, Publications of the Astronomical Society of Japan, 60, S153

\bibitem[{Bell(1978)}]{bell1978acceleration}
Bell, A. 1978, Monthly Notices of the Royal Astronomical Society, 182, 147

\bibitem[{Blandford \& Eichler(1987)}]{blandford1987particle}
Blandford, R., \& Eichler, D. 1987, Physics Reports, 154, 1

\bibitem[{Bohdan {et~al.}(2021)Bohdan, Pohl, Niemiec, Morris, Matsumoto, Amano, Hoshino, \& Sulaiman}]{bohdan2021magnetic}
Bohdan, A., Pohl, M., Niemiec, J., {et~al.} 2021, Physical review letters, 126, 095101

\bibitem[{Cotton {et~al.}(2024)Cotton, Kothes, Camilo, Chandra, Buchner, \& Nyamai}]{cotton2024meerkat}
Cotton, W., Kothes, R., Camilo, F., {et~al.} 2024, The Astrophysical Journal Supplement Series, 270, 21

\bibitem[{Dyer {et~al.}(2009)Dyer, Cornwell, \& Maddalena}]{dyer20091}
Dyer, K., Cornwell, T., \& Maddalena, R. 2009, The Astronomical Journal, 137, 2956

\bibitem[{Ellison {et~al.}(2010)Ellison, Patnaude, Slane, \& Raymond}]{ellison2010efficient}
Ellison, D.~C., Patnaude, D.~J., Slane, P., \& Raymond, J. 2010, The Astrophysical Journal, 712, 287

\bibitem[{Gardner \& Milne(1965)}]{gardner1965supernova}
Gardner, F., \& Milne, D. 1965, The Astronomical Journal, 70, 754

\bibitem[{Green(2001)}]{green2001}
Green, D.~A. 2001, Cambridge: Mullard Radio Astron: Obs

\bibitem[{Inoue \& Takahara(1996)}]{inoue1996electron}
Inoue, S., \& Takahara, F. 1996, Astrophysical Journal v. 463, p. 555, 463, 555

\bibitem[{Inoue {et~al.}(2011)Inoue, Yamazaki, Inutsuka, \& Fukui}]{inoue2011toward}
Inoue, T., Yamazaki, R., Inutsuka, S.-i., \& Fukui, Y. 2011, The Astrophysical Journal, 744, 71

\bibitem[{Jonas(2009)}]{jonas2009meerkat}
Jonas, J.~L. 2009, Proceedings of the IEEE, 97, 1522

\bibitem[{Kalemci {et~al.}(2006)Kalemci, Reynolds, Boggs, Lund, Chenevez, Renaud, \& Rho}]{kalemci2006x}
Kalemci, E., Reynolds, S.~P., Boggs, S.~E., {et~al.} 2006, The Astrophysical Journal, 644, 274

\bibitem[{Kataoka {et~al.}(1999)Kataoka, Mattox, Quinn, Kubo, Makino, Takahashi, Inoue, Hartman, Madejski, Sreekumar, {et~al.}}]{kataoka1999high}
Kataoka, J., Mattox, J., Quinn, J., {et~al.} 1999, The Astrophysical Journal, 514, 138

\bibitem[{Korreck {et~al.}(2004)Korreck, Raymond, Zurbuchen, \& Ghavamian}]{korreck2004far}
Korreck, K., Raymond, J., Zurbuchen, T., \& Ghavamian, P. 2004, The Astrophysical Journal, 615, 280

\bibitem[{Koyama {et~al.}(1995)Koyama, Petre, Gotthelf, Hwang, Matsuura, Ozaki, \& Holt}]{koyama1995evidence}
Koyama, K., Petre, R., Gotthelf, E., {et~al.} 1995, Nature, 378, 255

\bibitem[{Ksenofontov {et~al.}(2005)Ksenofontov, Berezhko, \& V{\"o}lk}]{ksenofontov2005dependence}
Ksenofontov, L.~T., Berezhko, E., \& V{\"o}lk, H. 2005, Astronomy \& Astrophysics, 443, 973

\bibitem[{Kundu(1970)}]{kundu1970brightness}
Kundu, M. 1970, Astrophysical Journal, vol. 162, p. 17, 162, 17

\bibitem[{Longair(1994)}]{Longair}
Longair, M.~S. 1994, {High Energy Astrophysics} (Cambridge University Press), \dodoi{10.1017/CBO9781139170505}

\bibitem[{Malkov \& Drury(2001)}]{malkov2001nonlinear}
Malkov, M., \& Drury, L.~O. 2001, Reports on Progress in Physics, 64, 429

\bibitem[{Miceli {et~al.}(2014)Miceli, Acero, Dubner, Decourchelle, Orlando, \& Bocchino}]{miceli2014shock}
Miceli, M., Acero, F., Dubner, G., {et~al.} 2014, The Astrophysical Journal Letters, 782, L33

\bibitem[{Milne(1971)}]{milne1971supernova}
Milne, D. 1971, Australian Journal of Physics, vol. 24, p. 757, 24, 757

\bibitem[{Milne \& Dickel(1975)}]{milne19755}
Milne, D., \& Dickel, J. 1975, Australian Journal of Physics, 28, 209

\bibitem[{Petruk {et~al.}(2009)Petruk, Dubner, Castelletti, Bocchino, Iakubovskyi, Kirsch, Miceli, Orlando, \& Telezhinsky}]{petruk2009aspect}
Petruk, O., Dubner, G., Castelletti, G., {et~al.} 2009, Monthly Notices of the Royal Astronomical Society, 393, 1034

\bibitem[{Reynolds \& Gilmore(1986)}]{reynolds1986radio}
Reynolds, S.~P., \& Gilmore, D.~M. 1986, Astronomical Journal (ISSN 0004-6256), vol. 92, Nov. 1986, p. 1138-1144., 92, 1138

\bibitem[{Reynoso {et~al.}(2013)Reynoso, Hughes, \& Moffett}]{reynoso2013radio}
Reynoso, E.~M., Hughes, J.~P., \& Moffett, D.~A. 2013, The Astronomical Journal, 145, 104

\bibitem[{Rothenflug {et~al.}(2004)Rothenflug, Ballet, Dubner, Giacani, Decourchelle, \& Ferrando}]{rothenflug2004geometry}
Rothenflug, R., Ballet, J., Dubner, G., {et~al.} 2004, Astronomy \& Astrophysics, 425, 121

\bibitem[{Rybicki \& Lightman(1985)}]{Rybicki:847173}
Rybicki, G.~B., \& Lightman, A.~P. 1985, {Radiative Processes in Astrophysics} (New York, NY: Wiley), \dodoi{10.1002/9783527618170}

\bibitem[{Saha {et~al.}(2014)Saha, Ergin, Majumdar, Bozkurt, \& Ercan}]{saha2014origin}
Saha, L., Ergin, T., Majumdar, P., Bozkurt, M., \& Ercan, E. 2014, Astronomy \& Astrophysics, 563, A88

\bibitem[{Uchiyama \& Aharonian(2008)}]{uchiyama2008fast}
Uchiyama, Y., \& Aharonian, F.~A. 2008, The Astrophysical Journal, 677, L105

\bibitem[{Uchiyama {et~al.}(2007)Uchiyama, Aharonian, Tanaka, Takahashi, \& Maeda}]{uchiyama2007extremely}
Uchiyama, Y., Aharonian, F.~A., Tanaka, T., Takahashi, T., \& Maeda, Y. 2007, Nature, 449, 576

\bibitem[{Winkler {et~al.}(2003)Winkler, Gupta, \& Long}]{winkler2003sn}
Winkler, P.~F., Gupta, G., \& Long, K.~S. 2003, The Astrophysical Journal, 585, 324

\bibitem[{Xing {et~al.}(2019)Xing, Wang, Zhang, \& Chen}]{xing2019fermi}
Xing, Y., Wang, Z., Zhang, X., \& Chen, Y. 2019, Publications of the Astronomical Society of Japan, 71, 77

\bibitem[{Zhou {et~al.}(2023)Zhou, Prokhorov, Ferrazzoli, Yang, Slane, Vink, Silvestri, Bucciantini, Reynoso, Moffett, {et~al.}}]{zhou2023magnetic}
Zhou, P., Prokhorov, D., Ferrazzoli, R., {et~al.} 2023, The Astrophysical Journal, 957, 55

\end{thebibliography}
\bibliographystyle{aasjournal}

\end{document}